\documentclass[traditabstract]{aa} 

\usepackage{times,epsfig,amssymb,amsmath,natbib}

\newcommand{\xmm}{{\it XMM-Newton}}
\newcommand{\planck}{{\it Planck}}

\usepackage{placeins}

 \usepackage{url}

\usepackage{float}
\usepackage[toc,page]{appendix}
\usepackage{tabularx}
\usepackage{braket}
	
\usepackage{natbib,twoopt}
\usepackage{verbatim}
\usepackage{amsmath}

\usepackage[]{hyperref}
\hypersetup{pdftex,colorlinks=true,allcolors=blue}
\usepackage{hypcap}
\usepackage{txfonts}
\usepackage{color}

\begin{document}

\title{The polytropic state of the intracluster medium \\ in the X-COP cluster sample}
\titlerunning{Polytropic ICM in X-COP}
\mail{vittorio.ghirardini@cfa.harvard.edu}

\author{V. Ghirardini\inst{1,2,3}\thanks{e-mail: \href{mailto:vittorio.ghirardini@cfa.harvard.edu}{\tt vittorio.ghirardini@cfa.harvard.edu}} \and S. Ettori\inst{2,4}\and D. Eckert\inst{5,6}\and S. Molendi\inst{7}}
\authorrunning{V. Ghirardini et al.}

\institute{
Center for Astrophysics | Harvard \& Smithsonian, 60 Garden Street, Cambridge, MA 02138, USA 
  \and INAF, Osservatorio di Astrofisica e Scienza dello Spazio, via Piero Gobetti 93/3, 40129 Bologna, Italy
 \and   Dipartimento di Fisica e Astronomia Universit\`a di Bologna, via Piero Gobetti, 93/2, 40129 Bologna, Italy 
 \and INFN, Sezione di Bologna, viale Berti Pichat 6/2, I-40127 Bologna, Italy
 \and Department of Astronomy, University of Geneva, Ch. d’Ecogia 16, CH-1290 Versoix, Switzerland
 \and Max-Planck-Institut f\"ur Extraterrestrische Physik, Giessenbachstrasse 1, 85748 Garching, Germany
 \and INAF - IASF Milano, via Bassini 15, I-20133 Milano, Italy
}

\mail{vittorio.ghirardini2@unibo.it}
\abstract
{}
{In this work, we investigate the relation between the radially-resolved thermodynamic quantities of the intracluster medium in the X-COP cluster sample, aiming to assess the stratification properties of the ICM.
}
{
We model the relations between radius, gas temperature, density and pressure using a combination of power-laws, also evaluating 
the intrinsic scatter in these relations.
}
{
We show that the gas pressure is remarkably well correlated to the density, with very small scatter. 
Also the temperature correlates with gas density with similar scatter.
The slopes of these relations have values that show a clear transition from the inner cluster regions to the outskirts. 
This transition occurs at the radius $r_t = 0.19 (\pm 0.04) R_{500}$ and electron density $n_t$ = $(1.91 \pm 0.21) \cdot 10^{-3} \text{cm}^{-3} E^2(z)$. 
We find that above $0.2 R_{500}$ the radial thermodynamic profiles are accurately reproduced by a well defined and physically motivated framework, 
where the dark matter follows the NFW potential and the gas is represented by a polytropic equation of state.
By modelling the gas temperature dependence upon both the gas density and radius, we propose a new method to reconstruct the hydrostatic mass profile based only on the quite inexpensive measurement of the gas density profile.
}
{}


\maketitle 

\section{Introduction}
The collapse of pristine gas into the dark matter gravitational potential of galaxy clusters heats up 
this optically-thin fully-ionized plasma at typical density of $10^{-3}-10^{-5}$ particles per cm$^3$ 
through compression and shock to very high temperature of 10$^{7}$--10$^8$ K, making this plasma an efficient source both in X-rays through Bremsstrahlung emission and at mm-wavelengths through the inverse-Compton scattering of the photons of the cosmic microwave background on its energetic electrons.

This hot intracluster medium (ICM) can be considered an almost perfect gas, with a ratio between specific heats, at constant pressure and at constant volume, of 5/3 (the adiabatic index). However both observations and simulations find that the ICM is effectively described by a polytropic equation $P_e = K n_e^{\Gamma}$. This is just an effective description of the global structure of the ICM, and does not represent how the gas pressure is affected by compression or expansion. In other words, $\Gamma$ is here considered not as the adiabatic index but as an {\it effective polytropic index}.

Simulations find that the polytropic equation generally provides a good description of the outer parts 
of the ICM, with $\Gamma$ in the range of 1.1 -- 1.3 \citep{Komatsu+01,Komatsu+02,Ascasibar+03,Ostriker+05,capelo+12}. 
Similarly, observations have measured this effective polytropic index, finding values very close to the predicted one \citep[][ measuring 1.24, 1.24, and 1.21 respectively]{Markevitch+98,Sanderson+03,eckert15}

An effective polytropic index is generally measured by fitting the gas pressure versus the density 
over a large radial range \citep[e.g. 0.1 -- 2 $R_{500}$\footnote{$R_{500}$ is defined as the 
radius within which the mean density is 500 times the critical density of the universe} in][]{eckert15}, 
thus providing poor constraints on how $\Gamma$ varies with radius, and therefore limited 
information on how the accretion history of clusters is occurring \citep[e.g.][]{Ascasibar+06,shi+16}.

{
In this paper, we make us of the unique dataset provided by the \emph{XMM-Newton} Cluster Outskirts Project \citep[X-COP, see ][]{xcop} to study the relations between the radial profiles of the gas density, temperature and pressure.

\begin{figure*}[ht]
\includegraphics[width=0.5\textwidth]{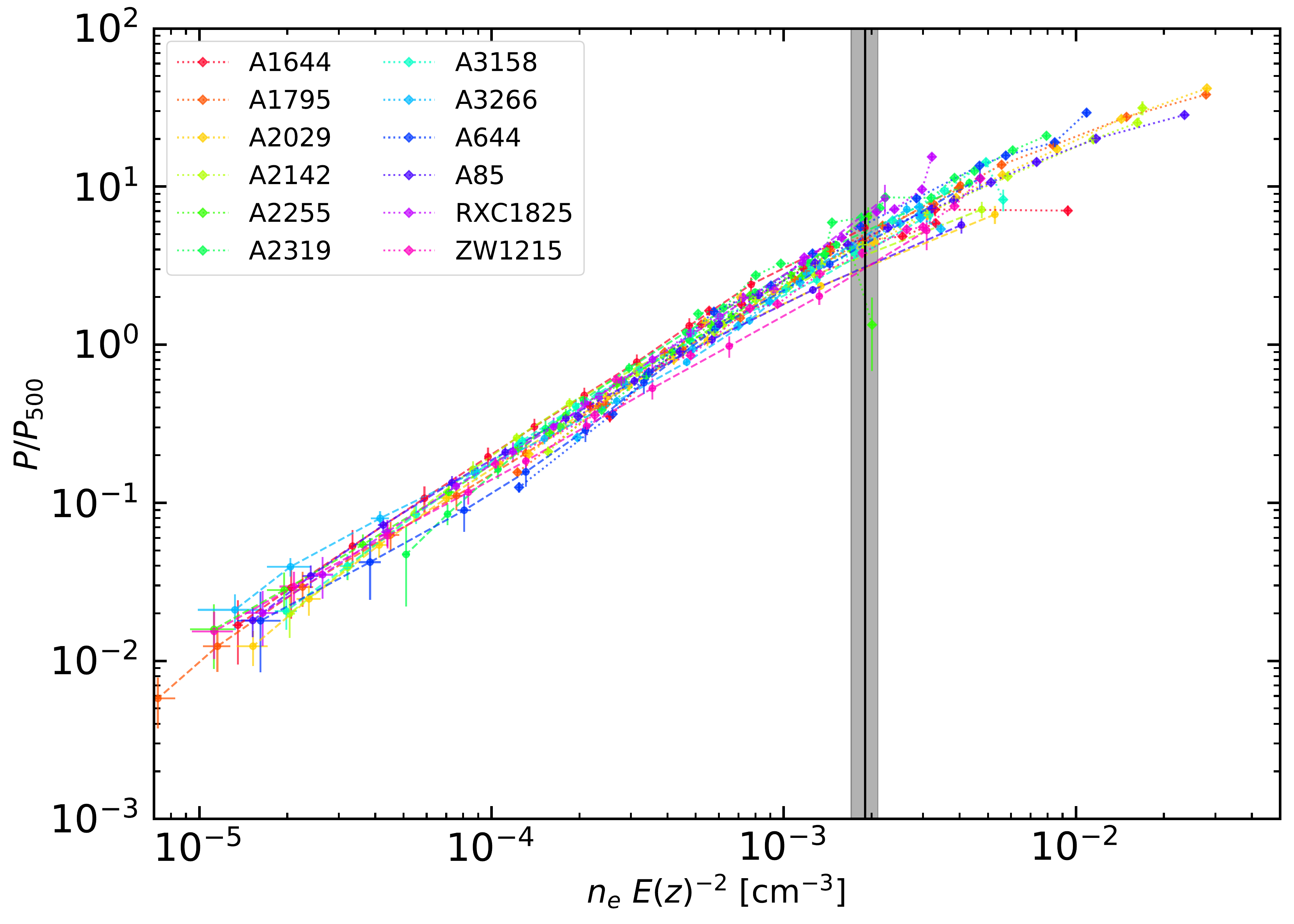}~
\includegraphics[width=0.5\textwidth]{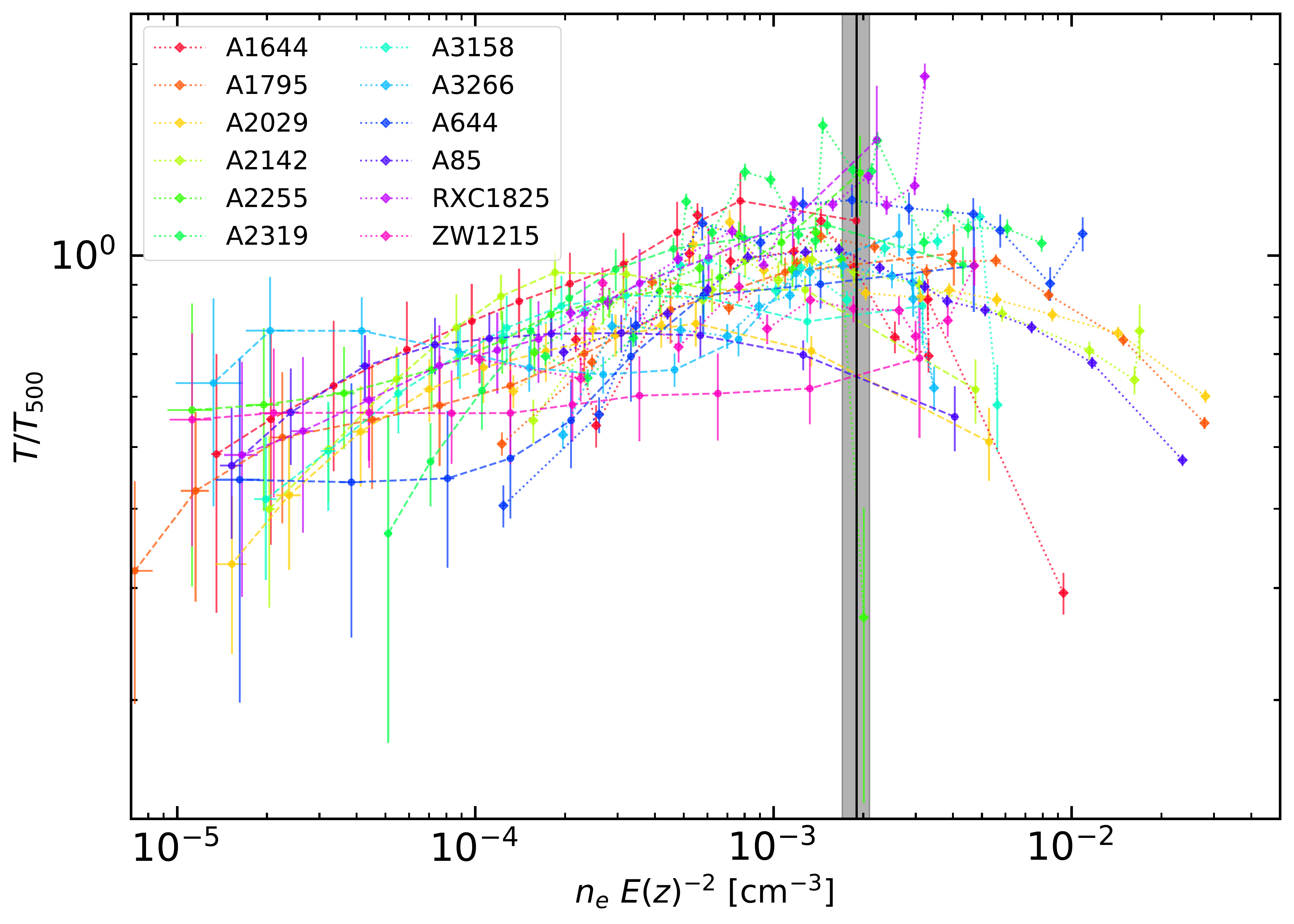}
\caption{(Left) Rescaled gas pressure profiles for the X-COP objects compared with the values of the electron density measured at the same radius.
(Right) Rescaled gas temperature profiles versus the electron density measured at the same radius.
{ The vertical lines indicate the location (with relative uncertainty) of the transition between core and outskirts discussed in Sect.~\ref{sec:2d}.}
}
\label{fig:Pn}
\end{figure*}

X-COP is a large \emph{XMM-Newton} program which aims at advancing significantly our knowledge on the physical conditions in the outskirts of galaxy clusters, by mapping the X-ray emission out to $\sim 2 R_{500}$ in 12 objects selected according to the following criteria:
S/N $> 12$ in the 1$^{\rm st}$ \emph{Planck} catalog \citep{PSZ1}; 
apparent size $\theta_{500} > 10$ arcmin; redshift in the range $0.04 < z < 0.1$;
Galactic $N_{\rm H} < 10^{21} \textrm{cm}^{-2}$. 
{
Although the targets satisfying the selection but with evident on-going mergers were excluded from the X-COP sample (and will be part of a recently approved AO18 \emph{XMM-Newton} program, PI: Ghirardini),
this sample includes mainly non-cool-core systems (8 out of 12), accordingly to their central entropy value \citep[see][]{xcop}, with hydrostatic masses $M_{500}$ in the range between $4\cdot10^{14} M_{\odot}$ and $10^{15} M_{\odot}$ \citep{ettori+18}.
}

The sample properties are extensively described in \cite{xcop} and \cite{ghirardini18_b}. 

} 
In particular, we present in \cite{ghirardini18_b}  
{ the universal behaviour of the radial profiles of the thermodynamic quantities spatially resolved in the X-COP objects,}
while in \cite{ettori+18}, we demonstrate that the Navarro–Frenk–White \citep[NFW,][]{nfw+97} model
provides the best representation of the gravitational potential for the same objects.

In this work, we study the relations between the radial profiles of the ICM 
pressure, density, and temperature out to 2 $\times R_{500}$, 
in order to { find how the ICM is stratified and the effective polytropic index varies as function of radius in the X-COP sample. This allows us to provide an effective description of the radial profiles of the ICM thermodynamical quantities both as a benchmark for hydrodynamical simulations and as a tool to make use of these correlations to infer other fundamental quantities, like the gravitating mass.
}

The paper is organized as follows. 
In Sect.~\ref{sec:2d}, we study { the relations between the radial profiles of the gas pressure, density and temperature}. 
We introduce the NFW polytropic model in Sect.~\ref{sec:poly},
showing how it provides a very good description of the intracluster medium outside the core.
{ In Sect.~\ref{sec:3d}, we model the temperature profile as function of radius and density profile, and apply this model to implement a very efficient tool to recover the hydrostatic mass profile when only the gas density profile is known.}
Finally, we draw our conclusions in Sect.~\ref{sec:conc}.
Throughout the paper, if not otherwise stated, we assume a flat concordance $\Lambda$CDM cosmology 
with $\Omega_m=0.3$, $\Omega_\Lambda=0.7$ and $H_{0}=70$ km s$^{-1}$ Mpc$^{-1}$, 
and use the Bayesian nested sampling algorithm MultiNest \citep{multinest} to constrain the best-fitting parameters. We define as ``$\log$'' the natural logarithm, while we indicate with ``$\log_{10}$'' the logarithm to the base 10.

\section{Relations between the radial profiles of the thermodynamic quantities in the ICM}
\label{sec:2d}

{
In Fig.~\ref{fig:Pn}, we show the measurements of the gas density, pressure 
and temperature spatially resolved in the X-COP clusters out to 2 $\times R_{500}$ \cite{ghirardini18_b}.
A tight power-law-like relation between { density and pressure} is present, 
with just a mild change in the slope in the high gas density regime.

To constrain the relation between the thermodynamic quantities, we implement a method similar 
to the piecewise power-law fitting procedure described in \cite{ghirardini18_b}.
We define 5 intervals in density (see Table~\ref{tab:Pn} and Table~\ref{tab:Tn}) and fit the following functional forms:
\begin{equation}
\frac{Q_e}{Q_{500}} = Q_0 \left[ n_e E(z)^{-2} \right]^{\Gamma_Q} \exp(\pm \sigma_{int,Q})
\label{eq:piece}
\end{equation}
where $Q$ represents either the temperature (T) or the pressure (P), and $Q_{500}$ is their representative value within the overdensity of 500\textrm{\footnote{ $P_{500}$ and $T_{500}$ are defined in Eq. (8) and (10), respectively, of \cite{ghirardini18_b}}}. 
For each bin in density, $Q_0$ is simply the normalization of each rescaled quantity, $\Gamma_Q$ is the logarithmic slope, and $\sigma_{int,Q}$ is the intrinsic scatter in the relation.

We find that the relation between the pressure and the density, or between the temperature and the density, is quite tight, with a scatter that has a mean value of about 0.15, and is always in the range 0.11--0.25, with the latter value measured only in the core, where the cluster population is characterized by a mix of systems with or without a cool core  \citep[see e.g.][]{cavagnolo+09}.

In the plane ``pressure versus density'', we notice two density regimes, one where the slope of the fitting power-law is smaller than 1, and another where this slope is bigger than 1 and approaches the value of 1.2, close to the predictions from numerical simulations \citep{capelo+12}.
For sake of completeness, we perform the fit of the rescaled pressure versus density 
in each X-COP object considering only the radial points above 0.2 $R_{500}$, 
and obtain values of $\Gamma$ between 1.1 and 1.3 (see Fig.~\ref{fig:fit_all}). 

\begin{figure}
\centering
\includegraphics[width=0.5\textwidth]{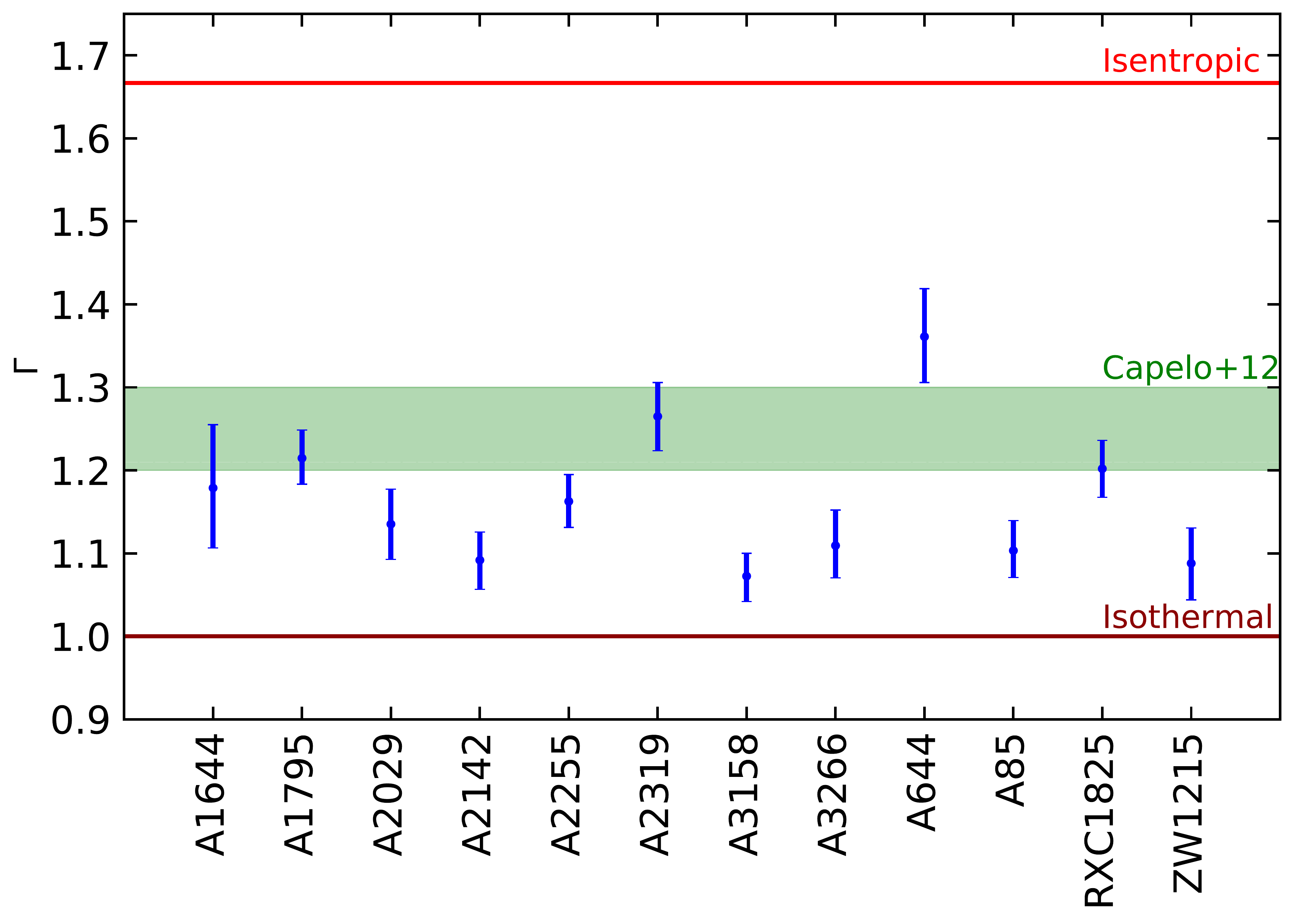}    
\caption{Results on the effective polytropic index $\Gamma$ obtained from a single powerlaw fit of the gas pressure versus the gas density on each X-COP cluster individually. { The values predicted from numerical simulations in \citep{capelo+12} are shown as horizontal green stripe.}}
\label{fig:fit_all}
\end{figure}

The analysis of ``temperature versus density'' confirms this trend.
In particular, the density value where this transition in the gradient happens 
locates the passage between the core region and the outskirts, 
allowing to define the core as the region where $\Gamma_P < 1$ and $\Gamma_T < 0$.
We can also constrain the value of the density where this transition occurs 
by using a broken power law to fit our data:
\begin{equation}
    \frac{Q_e}{Q_{500}}= 
\begin{cases}
    Q_0 \left[ n_e E(z)^{-2} \right]^{\Gamma_{Q,0}} \exp(\pm \sigma_{int,Q,0}),& \text{if } n_e E(z)^{-2} \leq n_t\\
    Q_1 \left[ n_e E(z)^{-2} \right]^{\Gamma_{Q,1}} \exp(\pm \sigma_{int,Q,1}),& \text{if } n_e E(z)^{-2} \geq n_t.
\end{cases}
\label{eq:transition}
\end{equation}
The continuity at the break location is forced by setting $Q_1 = Q_0 n_0^{\Gamma_{0}-\Gamma_{1}}$.
In the plane ``pressure versus density'', we find 
$n_t = (1.91 \pm 0.21) \cdot 10^{-3} \textrm{cm}^{-3}$,
$ \Gamma_{P,0} = 1.17 \pm 0.01$,
$ \Gamma_{P,1} = 0.77 \pm 0.04$,
$\log P_0 = 8.96 \pm 0.09$,
$\sigma_{int,P,0} = 0.15 \pm 0.01$,
and $\sigma_{int,P,1} = 0.24 \pm 0.01$. 
In the plane ``temperature versus density'', we obtain
$n_t = (1.93 \pm 0.23) \cdot 10^{-3} \textrm{cm}^{-3}$,
$ \Gamma_{T,0} = 0.17 \pm 0.01$,
$ \Gamma_{T,1} = -0.23 \pm 0.04$,
$\log T_0 = 1.12 \pm 0.09$,
$\sigma_{int,T,0} = 0.15 \pm 0.01$,
and $\sigma_{int,T,1} = 0.24 \pm 0.01$.

The same analysis can be done using bins in radius, instead of density.
The results of the piecewise analysis are displayed in Table~\ref{tab:Pn} and Table~\ref{tab:Tn} for pressure and temperature, respectively. They are very consistent with the ones from the previous analysis.

We can also use Eq.~\ref{eq:transition} and change the transition condition from $n_e E(z)^{-2} \geq n_t$ to $R/R_{500} \geq r_t$.
In the plane ``pressure versus density'', we find 
$r_t = (0.19 \pm 0.02)$,
$ \Gamma_{P,0} = 1.17 \pm 0.01$,
$ \Gamma_{P,1} = 0.78 \pm 0.04$,
$\sigma_{int,P,0} = 0.15 \pm 0.01$,
and $\sigma_{int,P,1} = 0.23 \pm 0.02$. 
In the plane ``temperature versus density'', we obtain
$r_t = (0.19 \pm 0.02)$,
$ \Gamma_{T,0} = 0.17 \pm 0.01$,
$ \Gamma_{T,1} = -0.21 \pm 0.04$,
$\sigma_{int,T,0} = 0.15 \pm 0.01$,
and $\sigma_{int,T,1} = 0.24 \pm 0.02$. 

We can thus conclude that a cluster core appear naturally from the analysis of these profiles,
and is defined as the region within $r_t = (0.19 \pm 0.02) R_{500}$ where the density is 
larger than $n_t = (1.9 \pm 0.2) \cdot 10^{-3} \textrm{cm}^{-3} E^2(z)$.
At electron density below this value, and radii larger than $r_t$, the ICM behaves very regularly and smooth at the level that a single power-law is able to describe the relations between pressure, density, temperature and radius.

\begin{table}
\caption{Results of the piecewise power-law fits on the gas pressure versus density in 
different (upper table) density and (lower table) radial ranges.
}
\centering
\setlength\tabcolsep{1.5pt} 
\begin{tabular}{c c c c c}
\hline
$n_{in} E(z)^{-2}$ & $n_{out} E(z)^{-2}$ & $\log(P_0)$ & $\Gamma_P$ & $\sigma_{int,P}$\\
\hline
7.22e-06 & 1.23e-04 & $ 8.92 \pm  0.45$ & $1.16 \pm 0.05$ & $0.11 \pm 0.02$\\
1.23e-04 & 3.55e-04 & $ 9.42 \pm  0.65$ & $1.23 \pm 0.08$ & $0.15 \pm 0.02$\\
3.56e-04 & 1.06e-03 & $ 8.80 \pm  0.63$ & $1.14 \pm 0.08$ & $0.15 \pm 0.02$\\
1.09e-03 & 2.64e-03 & $ 8.67 \pm  0.65$ & $1.13 \pm 0.10$ & $0.19 \pm 0.02$\\
2.85e-03 & 2.81e-02 & $ 6.47 \pm  0.29$ & $0.77 \pm 0.06$ & $0.25 \pm 0.03$\\ 
\hline 
$x_{in}$ & $x_{out}$ & $x_{average}$ & $\Gamma_P$ & $\sigma_{int,P}$ \\
\hline
0.013 & 0.112 & 0.058 &  $0.783 \pm 0.057$ & $0.26 \pm 0.03$\\ 
0.113 & 0.251 & 0.168 &  $0.844 \pm 0.071$ & $0.20 \pm 0.02$\\ 
0.263 & 0.451 & 0.327 &  $1.036 \pm 0.084$ & $0.17 \pm 0.02$\\ 
0.451 & 0.712 & 0.565 &  $1.293 \pm 0.061$ & $0.14 \pm 0.02$\\ 
0.720 & 1.148 & 0.885 &  $1.216 \pm 0.076$ & $0.16 \pm 0.02$\\ 
1.162 & 2.645 & 1.729 &  $1.203 \pm 0.046$ & $0.03 \pm 0.03$\\ 
\hline
\end{tabular}
\label{tab:Pn} 
\end{table}

\begin{table}
\caption{Same as Table~\ref{tab:Pn} but for the analysis with the gas temperature. 
}
\centering
\setlength\tabcolsep{1.5pt} 
\begin{tabular}{c c c c c}
\hline
$n_{in} E(z)^{-2}$ & $n_{out} E(z)^{-2}$ & $\log(T_0)$ & $\Gamma_T$ & $\sigma_{int,T}$\\
\hline
7.22e-06 & 1.23e-04 & $ 1.12 \pm  0.44$ & $0.16 \pm 0.05$ & $0.11 \pm 0.02$\\
1.23e-04 & 3.55e-04 & $ 1.65 \pm  0.64$ & $0.23 \pm 0.08$ & $0.15 \pm 0.02$\\
3.56e-04 & 1.06e-03 & $ 0.99 \pm  0.59$ & $0.14 \pm 0.08$ & $0.15 \pm 0.02$\\
1.09e-03 & 2.64e-03 & $ 0.76 \pm  0.67$ & $0.12 \pm 0.10$ & $0.19 \pm 0.02$\\
2.85e-03 & 2.81e-02 & $-1.35 \pm  0.28$ & $-0.23 \pm 0.05$ & $0.25 \pm 0.03$\\
\\ 
\hline 
$x_{in}$ & $x_{out}$ & $x_{average}$ & $\Gamma_T$ & $\sigma_{int,T}$ \\
\hline
0.013 & 0.112 & 0.058 &  $-0.211 \pm 0.058$ & $0.27 \pm 0.03$\\ 
0.113 & 0.251 & 0.168 &  $-0.129 \pm 0.068$ & $0.19 \pm 0.02$\\ 
0.263 & 0.451 & 0.327 &  $0.066 \pm 0.084$ & $0.17 \pm 0.02$\\ 
0.451 & 0.712 & 0.565 &  $0.288 \pm 0.062$ & $0.14 \pm 0.02$\\ 
0.720 & 1.148 & 0.885 &  $0.174 \pm 0.073$ & $0.16 \pm 0.02$\\ 
1.162 & 2.645 & 1.729 &  $0.187 \pm 0.045$ & $0.03 \pm 0.03$\\ 
\hline
\end{tabular}
\label{tab:Tn} 
\end{table}
}

\begin{figure*}[h!]
\centering
\includegraphics[width=0.45\textwidth]{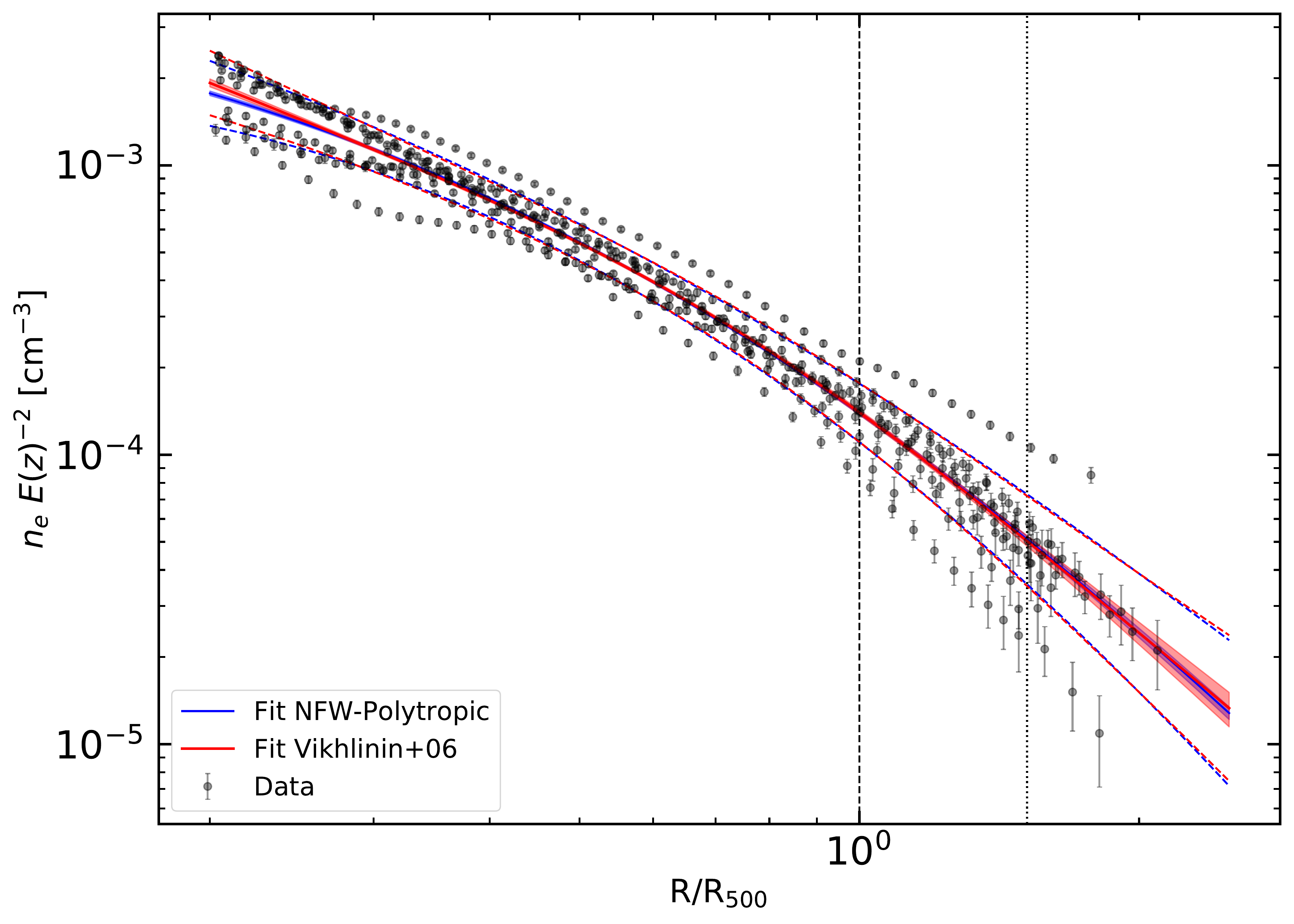}~
\includegraphics[width=0.45\textwidth]{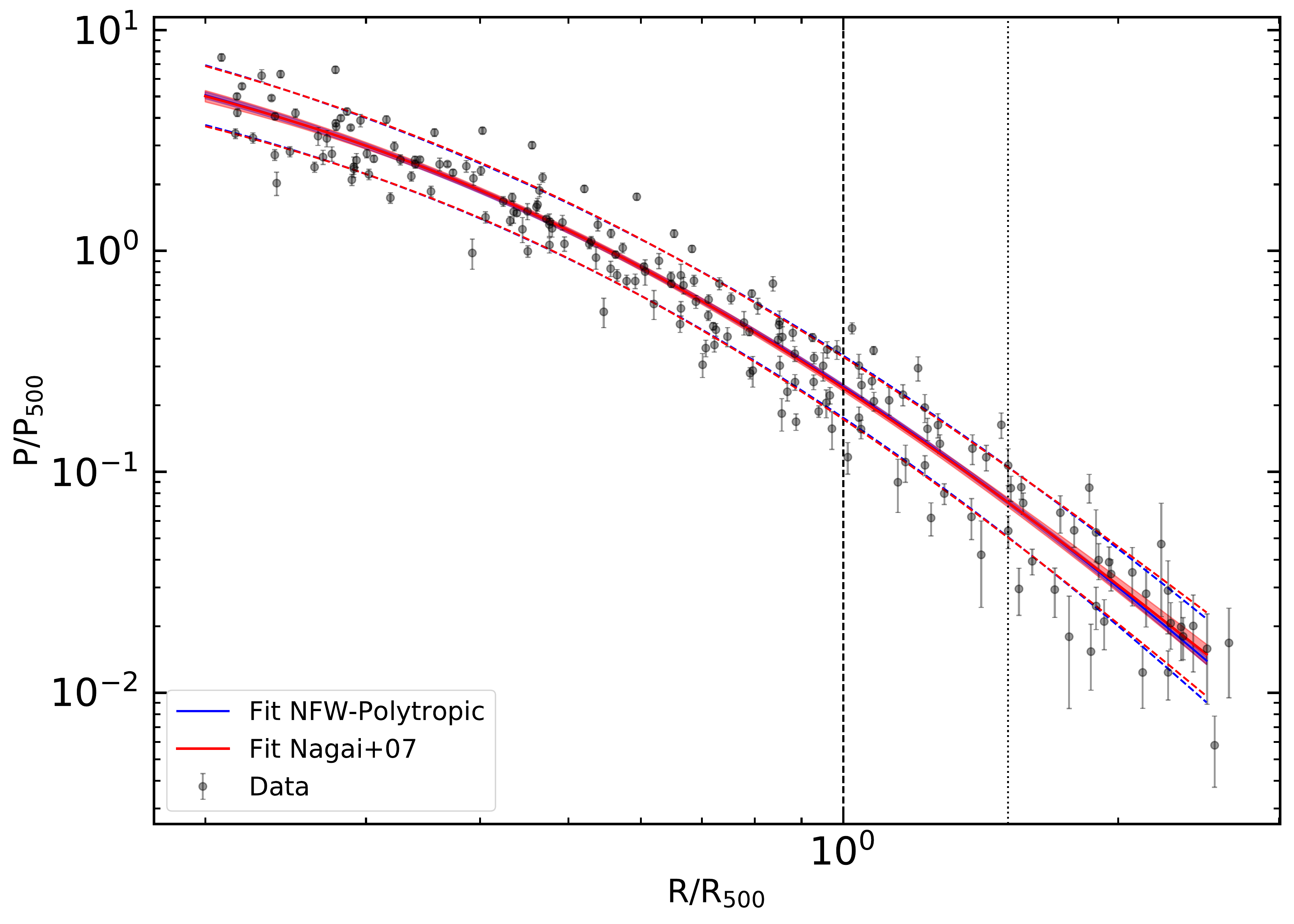}
\includegraphics[width=0.45\textwidth]{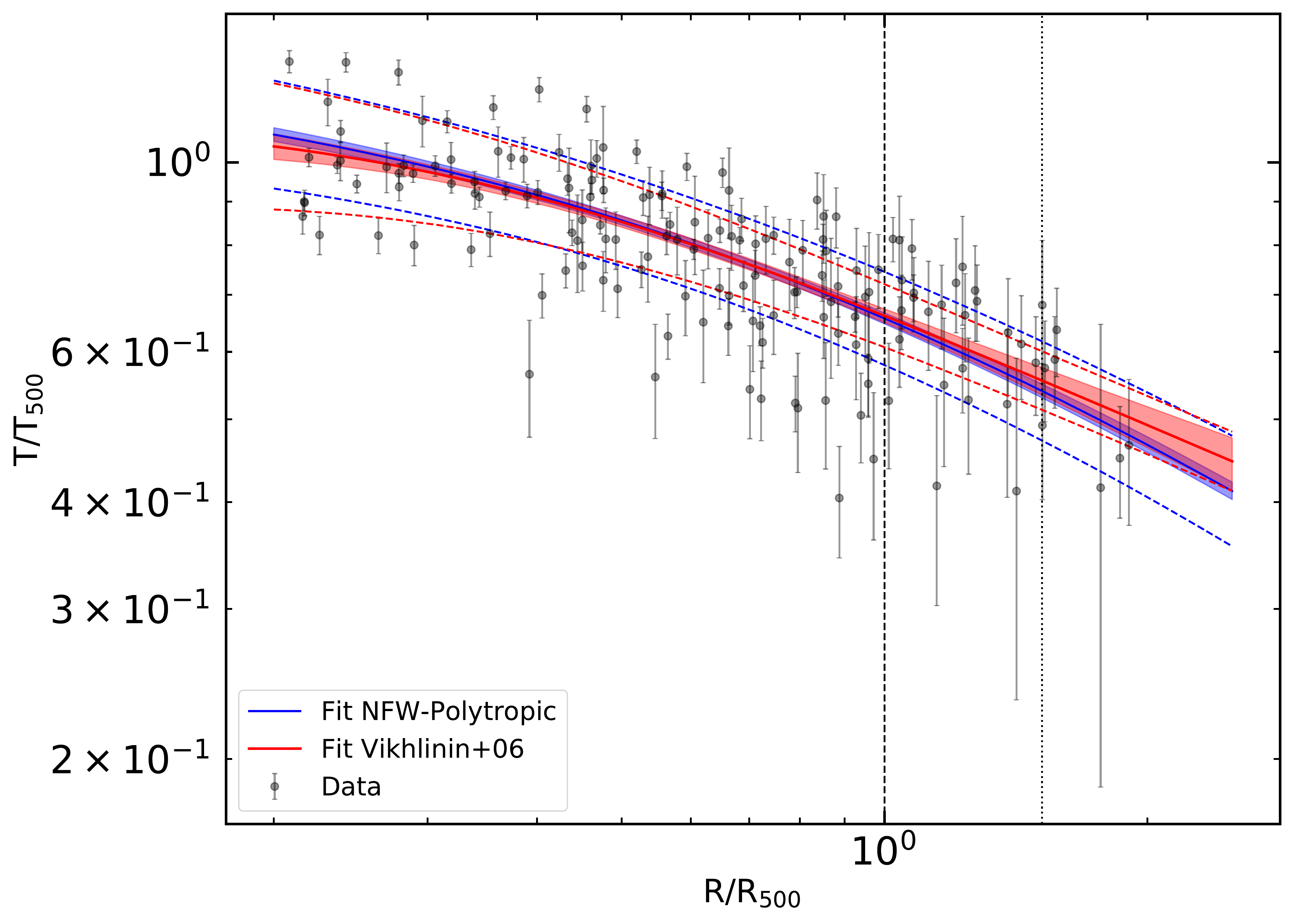}~
\includegraphics[width=0.45\textwidth]{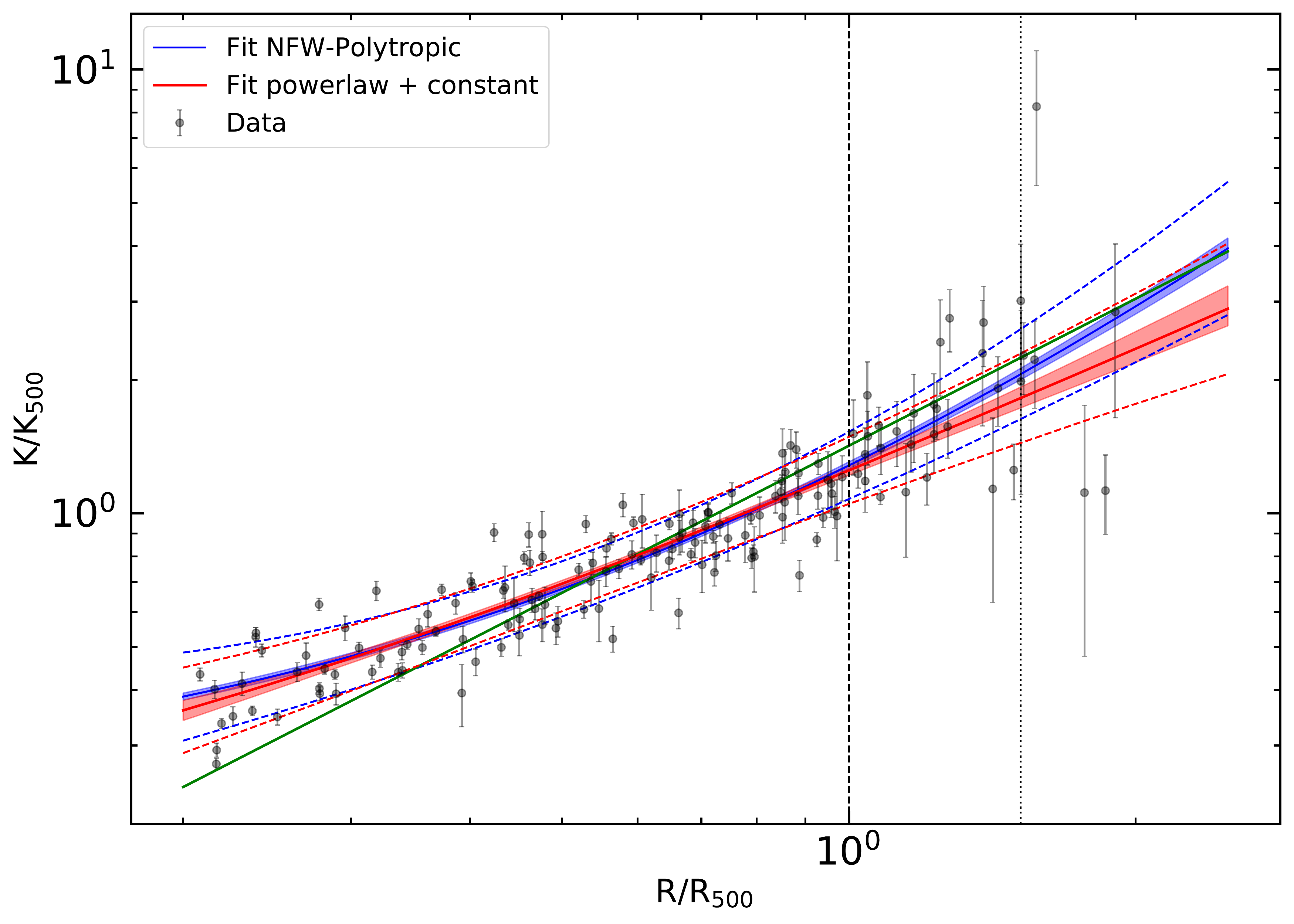}
\caption{ { Profiles of the (top) electron density and pressure, (bottom) temperature and entropy with, overplotted in blue, the 
joint fit obtained using the NFW-polytropic functional forms 
relating together the parameters $c_{500}$ and $\Gamma$ (see Table~\ref{tab:fit} and Sect.~\ref{sec:poly}), and, in red, 
the functional form used in \cite{ghirardini18_b}.
The shaded area around the functional form represents the 1$\sigma$ confidence level on the best-fit, while the dashed lines represents the scatter around the best fit.}
{The vertical dashed and dotted line represent the location of $R_{500}$ and $R_{200}$, respectively.}}
\label{fig:fit_compare}
\end{figure*}

\section{A polytropic NFW model for the gas}
\label{sec:poly}

The polytropic equation reads the following relation between the gas (electron) pressure and 
the gas (electron) density
\begin{equation}
P_e = \, C  \, n_e^{\Gamma}
\label{eq:poly}
\end{equation}
where $C$ is a constant, and $\Gamma$, also a constant, is the effective polytropic index.

\cite{bulbul+10} \citep[see also][]{Komatsu+01,Komatsu+02} showed that an ICM described by Eq.~\eqref{eq:poly} and in hydrostatic equilibrium with a gravitational potential modelled by NFW profile allows to write
\begin{equation}
n_e(x)^{\Gamma-1} \propto \frac{\log (1+c_{500} x)}{x},
\label{eq:nfw_poly}
\end{equation}
where $x = R/R_{500}$, { where for the NFW mass model $R_{500}=r_s c_{500}$}, with $r_s$ and $c_{500}$ are the NFW scale radius and concentration
(estimated within $R_{500}$), respectively.

In \cite{ettori+18}, we have shown that the NFW mass model is the best representation for the gravitational potential 
in the X-COP objects, as evaluated by computing the Bayesian evidence for several possible two-parameters mass models, with or without a core and with different slopes in the external parts. 
Also in the few cases where NFW is not the best fitting mass model, 
it does not show any statistically significant tension with the best fitting mass model.

{
Combining this evidence with the observed tight correlation between the gas density and the gas pressure 
at large radii (or in low density regime) that we discuss in the previous Section, 
}
allows us to model the stratification of the ICM with a {\it NFW-polytropic} profile 
\citep[see e.g. ][ with $\beta = 2$]{bulbul+10}:
\begin{equation}
E(z)^{-2} \, n_e(x) = n_{0} f(x)^\frac{1}{\Gamma-1},
\label{eq:nfw_func}
\end{equation}
where $x = r / R_{500}$ and $f(x) = \log (1+c_{500} x) / x$.
The correlated relations for pressure, temperature and entropy profile are then
$P(x) / P_{500} = P_{0} f(x)^{\Gamma /(\Gamma -1)}$,
$T(x) / T_{500} = T_0 f(x)$, 
$K(x) / K_{500} = K_0 f(x)^{(\Gamma -5/3)/(\Gamma-1)}$,
respectively, with $n_{0}$, $P_{0}$, $T_{0}$, and $K_{0}$ being the normalization factors, 
$c_{500}$ the NFW concentration, and $\Gamma$ the effective polytropic index. 

In principle, if these were just generic universal functional forms, 
the parameters do not have any physical meaning, and therefore their values are not expected 
to be the same in different thermodynamic properties.
However, if the physical assumptions behind this model are valid and robust, then the values 
of $c_{500}$ and $\Gamma$ should be the same for all the profiles of the thermodynamic quantities, 
and should represent the properties of the underlying mass distribution assumed to be in hydrostatic equilibrium.

To evaluate the performance of this functional forms, we consider only the radial range 
beyond $\sim 0.2 R_{500}$, where we have demonstrated that the assumption of constant $\Gamma$ 
is strictly valid.

First, we fit each thermodynamic quantities independently. 
We show the results of the fit in Table~\ref{tab:fit}.
We notice that $c_{500}$ and $\Gamma$ from the fits are compatible at the 1$\sigma$ level.
This indicates that the NFW-polytropic model is robust and consistent enough to characterize the outer regions of galaxy clusters.

Then, we proceed with a joint-fit, i.e. forcing $c_{500}$ and $\Gamma$ to be the same for all thermodynamic quantities, showing the best fitting results in the last row of Table~\ref{tab:fit}.
In Fig.~\ref{fig:fit_compare}, we compare the best-fit profiles with the functional forms 
described in \cite{ghirardini18_b}. 
We observe that the NFW-polytropic is a very good fit to the data, and very well consistent \citep[with differences in Bayesian Evidences smaller than 5;][]{jef61} 
with the fit performed with other functional functional forms adopted in the literature 
(e.g. \cite{vikhlini+06} for density and temperature; \cite{nagai+07} for pressure; \cite{cavagnolo+09} for entropy). 
On the other hand, the improvement is dramatic for what concerns both the physical interpretation of the 
best-fit parameters, and the simplification of the fitting procedure, considering the limited 
number of parameters in this new functional forms and the lack of any degeneracy among these parameters.

\begin{table}
\caption{\label{tab:fit} Best fitting parameters for the functional form described in Sect.~\ref{sec:poly}.
We define the following priors: 
$\mathcal{U}(-13,-9)$ on $\log(n_0)$; $\mathcal{U}(-5,-1)$ on $\log(P_0)$; $\mathcal{U}(-2,2)$ on $\log(T_0)$;
$\mathcal{U}(-4,4)$ on $\log(K_0)$; $\mathcal{U}(0,2)$ on $\log(c_{500})$; $\mathcal{U}(1,1.5)$ on $\Gamma$.
The column ``$\log(N)$'' show the best-fit normalization, with $N = n_0, P_0, T_0, K_0$ for density, pressure, temperature and entropy, respectively.
The normalizations in the ``Joint fit'' row refer to (from top to bottom)
density, pressure, temperature and entropy, respectively.
}
\centering
\begin{tabular}{ c c c c c }
\hline
Quantity & $\log(N)$ & $c_{500}$ & $\Gamma$ \\
\hline
\bf Density & $-10.3 \pm 0.4$ & $2.69 \pm 0.43$  &  $1.20 \pm 0.01$ \\ 
\bf Pressure & $-2.94 \pm 0.61$ & $2.61 \pm 0.55$  &  $1.19 \pm 0.02$ \\ 
\bf Temperature & $-0.60 \pm 0.10$ & $2.34 \pm 0.35$  \\ 
\bf Entropy & $0.92 \pm 0.34$ & $2.92 \pm 1.02$  &  $1.21 \pm 0.02$ \\
\\
{\bf Joint fit} & $-10.2 \pm 0.2$ & $2.64 \pm 0.24$ &  $1.19 \pm 0.02$ \\ 
  & $-2.99 \pm 0.26$ & \\
  & $-0.68 \pm 0.06$ & \\
 & $0.87 \pm 0.08$ &  \\
\hline
\end{tabular}
\end{table}

\begin{figure}[t]
\includegraphics[width=0.5\textwidth]{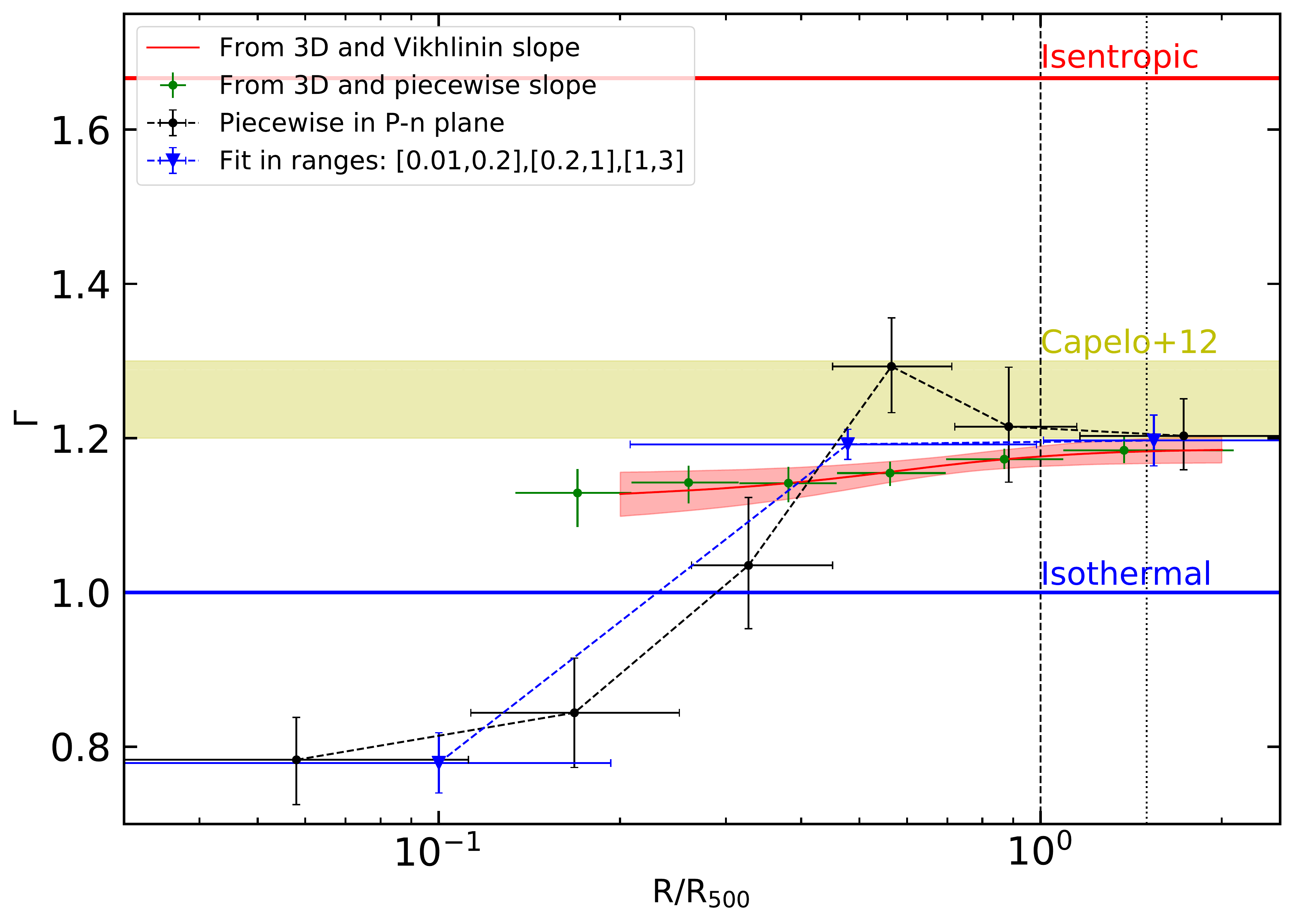}
\caption{{ Effective polytropic index as function of radius.
Black data points represent the results of the piecewise fit from Eq.~\ref{eq:piece}. 
Also the blue data points come from a piecewise analysis, but with radii chosen a posteriori.
Red curve and green points indicate the polytropic index estimated from our combined analysis 
(see Sect.~\ref{sec:3d}), the difference being in the slope of the density considered
(see text after Eq.~\ref{eq:gamma_Tr}).
The horizontal blue and red lines indicate the location of the isothermal and isentropic limits, respectively. 
{ The horizontal yellow stripe represents the value predicted by numerical simulations, as the green stripe in Fig.~\ref{fig:fit_all}.}
The vertical dashed and dotted line represent the location of $R_{500}$ and $R_{200}$, respectively.}
}
\label{fig:poly_r}
\end{figure}

{
\section{Combined analysis of the gas temperature, density and radius}
\label{sec:3d}

In general, the study of the profiles of the thermodynamic properties of the ICM is focused on each single quantity, independently \citep[see e.g.][]{ghirardini18_b}. However, connections between relative variations 
in e.g. gas density, temperature and pressure profiles are expected due to the nature of perfect gas of the ICM. 
In this section, we explore the possibility to analyze the combined variation of the gas density, temperature, and radius. 
We decide to treat the temperature instead of the pressure because it is not directly correlated with gas density. Moreover we consider only the region above 0.2 $R_{500}$, both because we are interested mainly in the outskirts where gravity dominates, and because our data are not suited to study the central regions of clusters, because of the large binning scheme adopted which allows to avoid effects due to the instrumental Point-Spread-Function of both \xmm\ and \planck.

\subsection{Linear model for T-n-R}
For this combined analysis, we use a simple model obtained from two power-laws fitted in the logarithmic space:
\begin{equation}
\log \left( \frac{T}{T_{500}} \right) = \log T_0 + \Gamma_0 \log \left[ n_e E^{-2}(z) \right] + \Gamma_{R} \log \left( \frac{R}{R_{500}} \right) \pm \sigma_{int},
\label{eq:3d_plane}
\end{equation}
where $T_0$, $\Gamma_0$, and $\Gamma_{R}$ are the free parameters. 
The distributions of the best fitting parameters are shown in Fig.~\ref{fig:plane} and are centered at $\log T_0 = 1.76 \pm 0.43$, $\Gamma_0 = 0.25 \pm 0.05$, and $\Gamma_R 0.14 \pm 0.09$.
It is worth noticing that the no null value of $\Gamma_{R}$ suggests that the model with a single relation between two quantities is not complex enough to fully characterize the thermodynamic quantities when high quality data are available. 
We also tested whether increasing the complexity of the fitted function (i.e. by using a quadratic or a cubic function) allows to improve the fit significantly by comparing the obtained bayesian evidences.
We find that Equation~\ref{eq:3d_plane} is not only the simplest model, but also the one with the highest bayesian evidence in reproducing the observed profiles in the X-COP sample. 

\begin{figure}[t]
\centering \includegraphics[width=0.5\textwidth]{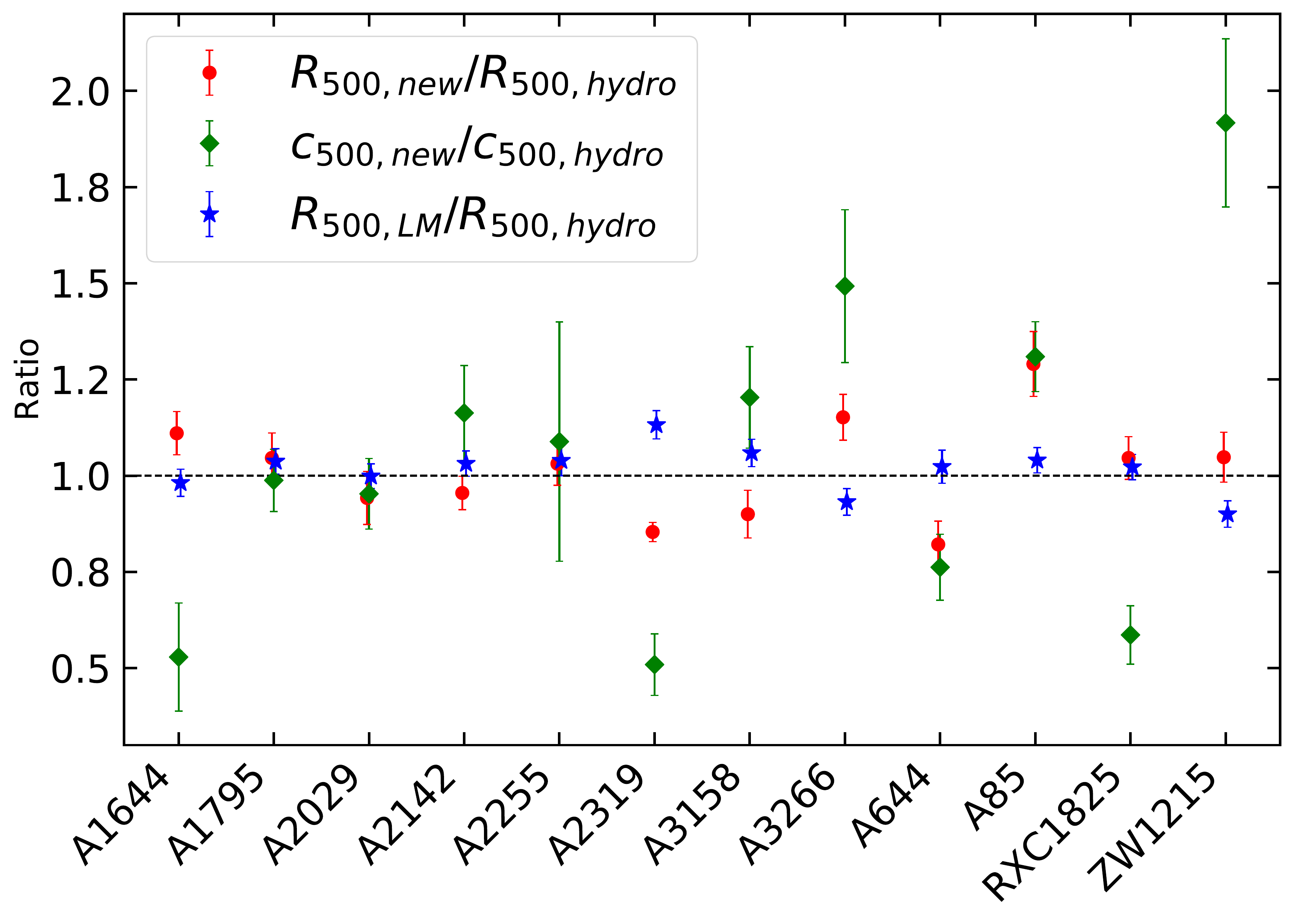}
\caption{
{
Ratios between the quantities recovered with the technique outlined in Sect.~\ref{sec:mass} ($R_{500, new}$, $c_{500, new}$) and the hydrostatic values ($R_{500, hydro}$, $c_{500, hydro}$). We show also the comparison with estimates from the $L-M$ relation ($R_{500, L-M}$).
} 
} \label{fig:comparison}
\end{figure}

An important implication of this model is that we can then write the effective polytropic index as
\begin{equation}
\Gamma (R) = \frac{d \log P}{d \log n_e} = \frac{d \log (n_e T)}{d \log n_e} = 
1 + \Gamma_0 +  \Gamma_{R}  \frac{\partial \log R}{\partial \log n_e} 
\label{eq:gamma_Tr}
\end{equation}
where we have used Eq.~\eqref{eq:3d_plane},
and $\frac{\partial \log R}{\partial \log n_e}$ has been calculated in \cite{ghirardini18_b} using two different techniques, (i) by measuring the slope in ranges using a piecewise power-law analysis, 
and (ii) by fitting a functional form to the entire density profile and then computing analytically the slope.
We show the resulting profile of the effective polytropic index in Fig.~\ref{fig:poly_r}. 
We notice that the small differences between this effective polytropic index, and the ones shown as data points can be attributed to the fact that the latter are missing a contribution coming from the last term in Eq.~\ref{eq:gamma_Tr}.

Finally, we note that the intrinsic scatter associated to this combined analysis, $\sigma_{int} = 0.15 \pm 0.01$, is smaller than all the scatters we have measured by studying the relation between each single thermodynamic quantity and the radius \citep[see Fig.~7 and Table~2 in ][]{ghirardini18_b}, indicating that, indeed, our combined analysis reproduces more closely the distribution of the observed quantities.

\subsection{The reconstruction of the total mass profile}
\label{sec:mass}

A direct, and very useful, application of the results of this combined analysis
is the reconstruction of the mass profile by using only the gas density profile, 
which is, typically, a very cheap measurement from X-ray observations.

Let us consider the hydrostatic equilibrium equation,
\begin{equation}
\frac{1}{\mu m_p n_e} \frac{dP_e}{dR} = - \frac{G M(<R)}{R^2},
\label{eq:HEE}
\end{equation}
where $\mu$ is the mean molecular mass weight and $m_p$ is the proton mass.
The term on the left hand side of Eq.~\ref{eq:HEE} depends only on the density profile, once a polytropic relation is adopted.
We can then reconstruct a model of the gas density profile given a mass model, like the NFW, which would depend only on the free parameters of such mass model. Tweaking the parameter of the mass model until the models of the gas density reproduces at the best the observed density profile allows to measure the parameters of such mass model, and finally the mass profile.
The detailed calculations which allow to build the model of the gas density are presented in Appendix~\ref{app:mass}.

We test this method using the results of our analysis of the X-COP data.
As discussed, we restrict our analysis to the outer cluster regions, over the same radial range 
where we have found that our combined analysis discussed in Sect.~\ref{sec:3d} 
produces the best representation of the thermodynamic quantities.
The exclusion of the core ($\la 0.2 R_{500}$) implies that we have a very weak leverage in constraining the concentration parameter of a NFW mass model.
Thus, we choose a gaussian prior on the concentration parameter, centered on the concentration--mass relation
provided by \cite{diemer18}\footnote{as implemented in the code {\tt COLOSSUS} \citep{Diemer+17}, with $\Omega_{\rm m}=0.3, \Omega_{\Lambda}=0.7, \sigma_8=0.8, H_0 = 70$ km s$^{-1}$ Mpc$^{-1}$}: $\log c_{500} = 0.885 -0.049 \log (M_{500}/5 \, 10^{14} M_{\odot})$.
An instrinsic scatter of $\sigma_{log_{10}(c_{500})} = 0.1$ \citep[from][]{Neto+07} is propagated through our analysis.
In Fig.~\ref{fig:comparison}, we show the comparison between the best-fit results on the NFW parameters 
recovered using both the method outlined here, that is based solely on the gas density profile, and 
the entire set of the observed radial profiles \citep[see][]{ghirardini18_b, ettori+18}. 
{
We obtain a very good agreement between these best-fit parameters, 
with a mean ratio of 1.02 (r.m.s 0.13) and 1.04 (0.40) for $R_{500}$ and $c_{500}$, respectively.
This is very similar to what we can reconstruct using a robust mass proxy like the 
X-ray bolometric luminosity. By adopting the scaling relation in \cite{pratt+09}, we recover a total mass with a corresponding $R_{500}$ that is, on average, 1.02 times the hydrostatic value, with a dispersion of 0.06.
The latter value of dispersion confirms that the use of robust mass proxy allows to recover the integrated quantities more accurately, whereas the technique we propose in this work provides a less accurate, although more complete, description of the entire mass profile.
}

\section{Conclusions}
\label{sec:conc}

We present the relation between the ICM thermodynamic quantities in the outskirts of 
12 SZ-selected galaxy clusters observed with \emph{XMM-Newton} and \emph{Planck} for the 
\emph{XMM-Newton} Cluster Outskirts Project \citep[X-COP,][]{xcop}.
Using the radial profiles recovered in \cite{ghirardini18_b}, we investigate 
the polytropic relation in the intracluster medium and 
the variation of the effective polytropic index $\Gamma$ as function of the radius.

Our main findings are:
\begin{itemize}
\item the gas pressure and density are tightly correlated, with scatter of 0.20--0.25 in the inner regions, and of about 0.10--0.15 in the outskirts; this is also valid when the relation between the gas temperature and density is considered;
\item the logarithmic slope in the relation between the gas pressure and density changes from the value of $\sim$~0.8, measured in the clusters' core, to the value of $\sim$~1.2 observed in the outskirts;
similarly, the logarithmic slope of the relation between the gas temperature and density changes from $-0.2$ in the core to $0.2$ in the outskirts;
\item this transition occurs where the electron density is lower than $1.9 (\pm 0.2) \cdot 10^{-3} E(z)^2 \text{cm}^{-3}$, 
or the radius is larger than $0.19 (\pm 0.02) R_{500}$;
\item relying on this tight single power-law relation between the gas pressure and density 
in the outskirts of the X-COP clusters, and using a NFW mass model, 
which provides the best description of the observed profiles of the thermodynamic quantities \citep[see][]{ettori+18}, 
we find that the {\it polytropic-NFW} model in Eq.~\ref{eq:nfw_func}, originally introduced by \cite{bulbul+10}, fits the observed quantities equally well to what obtained from other models available in the literature, with the great advantages that 
(i) it is physically motivated, (ii) it makes use of few, well constrained parameters, and (iii) these parameters that can be interpreted as physically interesting quantities;
\item beyond the core, the polytropic state of the ICM is well represented by a model (see Eq.\ref{eq:gamma_Tr}), where the gas temperature depends on both  the gas density and the radius with a well constrained scatter;
\item {
as described in Section~\ref{sec:mass} (and detailed in Sect.\ref{app:mass}), the tight relation between temperature, density, and radius allows to estimate the parameters of a hydrostatic mass profile relying solely on the gas density profile, that is a quiet inexpensive X-ray measurement, with an assumption on the concentration--mass relation needed to break further some degeneracy present in the method. 
}
\end{itemize}

Larger samples, covering lower mass regime and at higher redshifts than the ones investigated in the X-COP sample,
are needed to explore the robustness and the universality of our conclusions on the polytropic state of the ICM,
and will become available in the next future thanks to a dedicated \xmm\ Heritage program \footnote{\url{http://xmm-heritage.oas.inaf.it/}}.
{ In the meantime, our results represent the best benchmark available for hydrodynamical simulations on the radial behaviour of the effective polytropic index in the ICM.}

\begin{acknowledgements}
The research leading to these results has received funding from the European Union’s Horizon 2020 Programme under the AHEAD project (grant agreement n. 654215). S.E. acknowledges financial contribution from the contracts NARO15 ASI-INAF I/037/12/0, ASI 2015-046-R.0 and ASI-INAF n.2017-14-H.0.
\end{acknowledgements}

}
\bibliographystyle{aa} 
\bibliography{XCOP_poli} 

\begin{appendix}
\section{Radial profiles of gas pressure, temperature and density}
\label{sec:functional form fit}

\begin{figure}[h]
\includegraphics[width=0.5\textwidth]{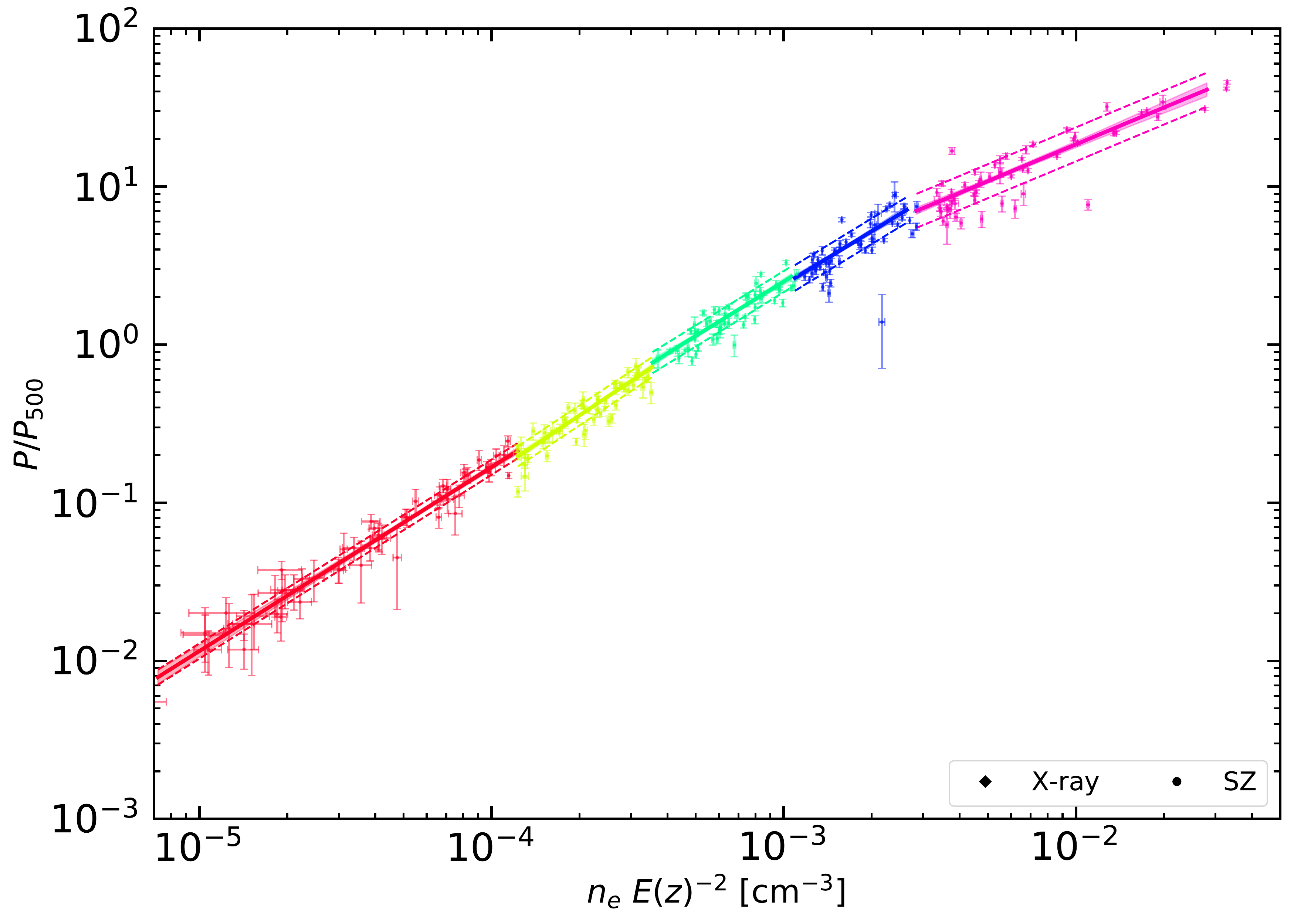}

\caption{ Piecewise power-law fit of the gas pressure versue density
(best-fit values quoted in Tab.~\ref{tab:Pn}).
Solid lines and contour indicate the best fitting result with statistical uncertainties; 
dashed lines indicate the intrinsic scatter in the distribution around the best fit. { The different colors correspond to different density range, where these ranges have been chosen to have the same number of points within them.}}
\label{fig:fit_full}
\end{figure}

\begin{figure}[h]
\includegraphics[width=0.45\textwidth]{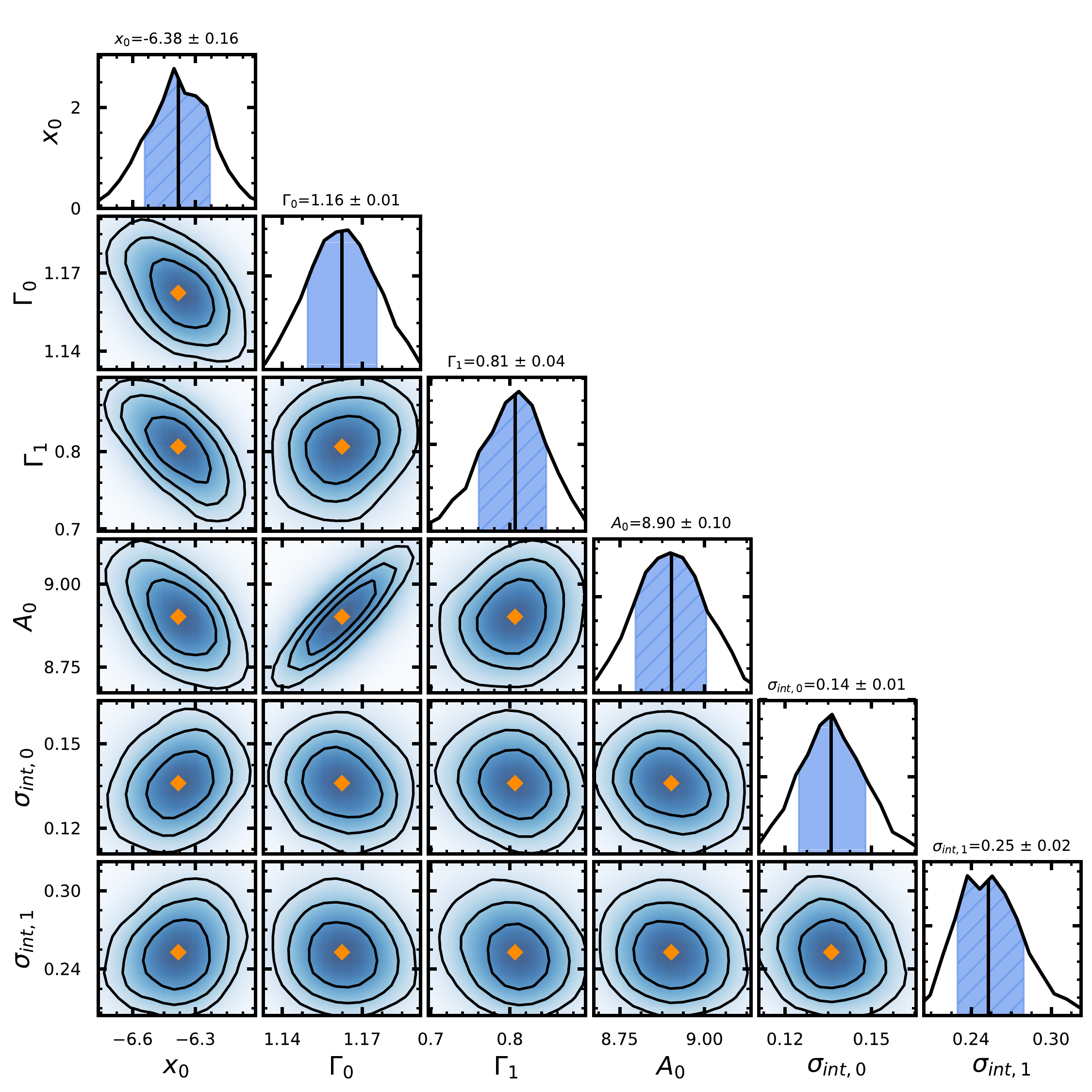}
\caption{Posterior distribution of the fitting parameters of the functional form in Eq.~\eqref{eq:transition}.}
\label{fig:fit_full2}
\end{figure}

\begin{figure}
\includegraphics[width=0.45\textwidth]{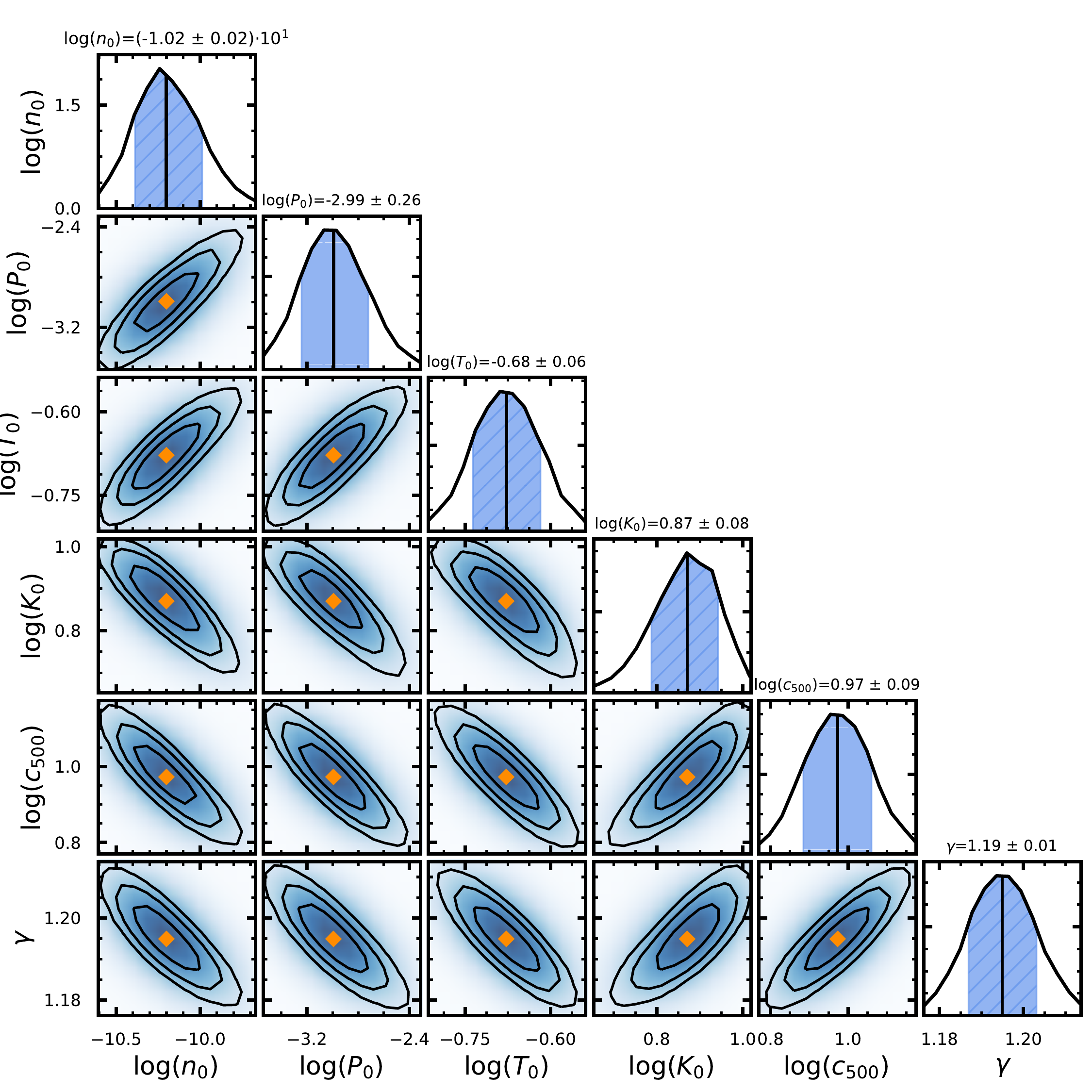}
\caption{Posteriors on the parameters of the the functional forms introduced in Section~\ref{sec:poly}, Eq.~\eqref{eq:nfw_func}, fitted jointly to the thermodynamic profiles.}
\label{fig:post_joint}
\end{figure}

\begin{figure}
    \centering
    \includegraphics[width=0.5\textwidth]{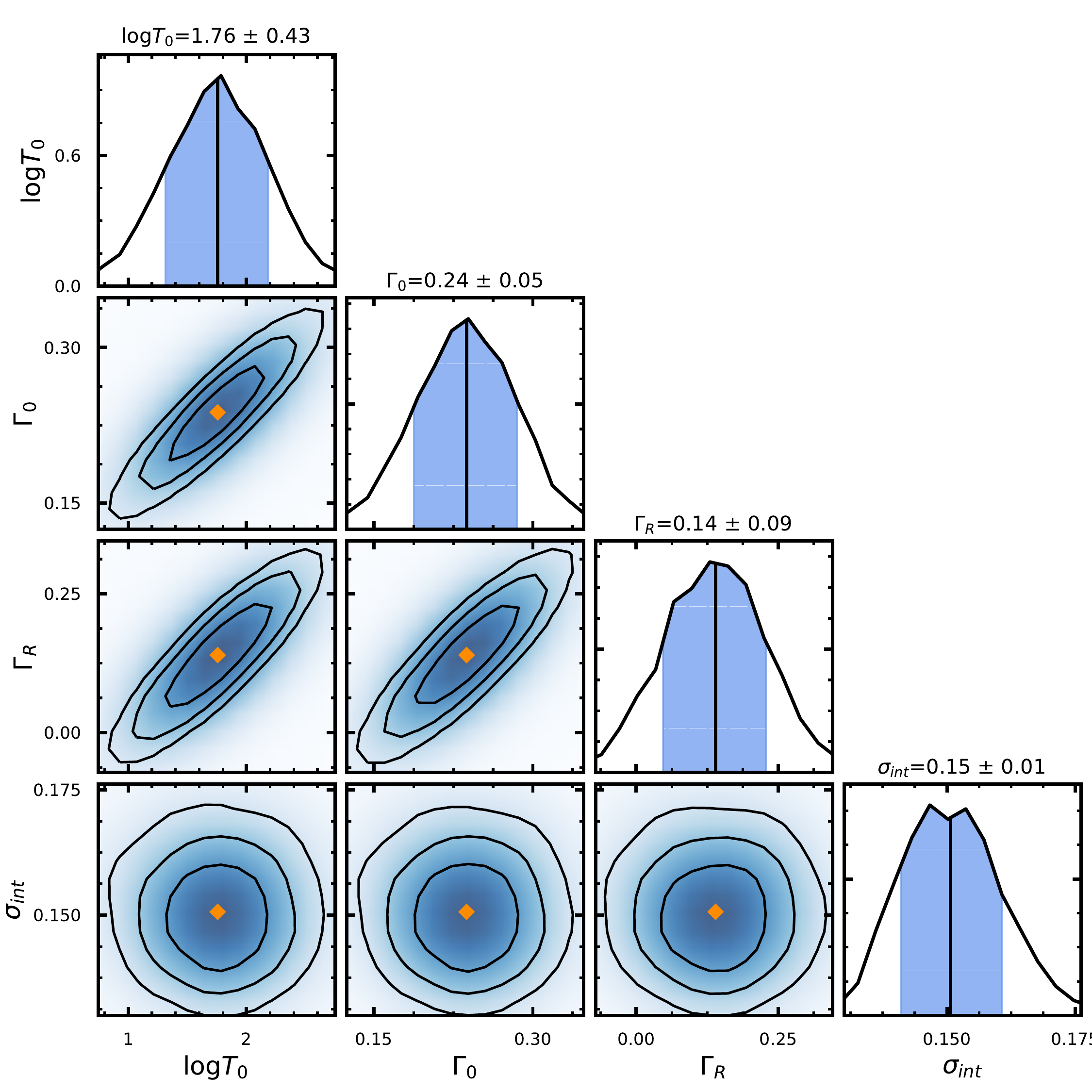}
    \caption{Best fitting parameters from the model described by Eq.~\ref{eq:3d_plane}.}
    \label{fig:plane}
\end{figure}

We present here a collection of plots that are relevant to 
(i) the analysis in the ``pressure vs density'' plane (Fig.~\ref{fig:fit_full} and Fig.~\ref{fig:fit_full2}), and
(ii) the modelization with a {\it polytropic NFW} functional form of the radial profile
of each thermodynamical quantity shown in Fig.~\ref{fig:fit_compare} and the results of the 
joint-fit with the same polytropic NFW functional form (Fig.~\ref{fig:post_joint}).

The posterior distribution of the parameters of the model described in Eq.~\ref{eq:3d_plane} and that combines the radius with the gas temperature and density profiles is shown in Fig.~\ref{fig:plane}.

\section{NFW polytropic model with $\Gamma (r)$}
\label{sec:gamma_r}

The functional forms presented in Section~\ref{sec:poly} can be also 
generalized by assuming that $\Gamma$ is a function of radius.
From the equation of the hydrostatic equilibrium (Eq.~\ref{eq:HEE}) 
with a NFW potential, we can write
\begin{equation}
\frac{1}{\rho} \frac{d \rho^\Gamma}{dx}  = \rho^{\Gamma-1} \frac{d}{dx} (\Gamma \log \rho)
= - \frac{G M(<r)}{r^2} = K_0 \frac{d}{dx} \left( \frac{\log(1+x)}{x} \right).
\label{eq:hee_gamma_r}
\end{equation}

Adopting the NFW polytropic modelling of the gas (Eq.~\ref{eq:nfw_poly};
$\rho_{\rm gas}^{\Gamma-1} = K_1 \, \log (1+x) / x$), 
we can expand the first term of the above equation:
\begin{equation}
\rho^{\Gamma-1} \frac{d}{dx} (\Gamma \log \rho) =
K_1 \frac{\log (1+x)}{x} \frac{d}{dx} \left[ \frac{\Gamma}{\Gamma-1} \log \left( K_1 \frac{\log(1+x)}{x} \right) \right].
\end{equation}
By using this expansion in Eq.~\eqref{eq:hee_gamma_r}, and moving the terms with $y = \log (1+x) / x$ on the same side, we can write:
\begin{equation}
\frac{K_0}{K_1} \frac{\frac{dy}{dx}}{y} = \frac{d}{dx} \left[ \frac{\Gamma}{\Gamma-1} 
\log \left( K_1 y \right) \right].
\end{equation}
Finally, converting from $\frac{1}{y}\frac{dy}{dx}$ to $\frac{d\log(y)}{dx}$ and  
integrating on both side we have:
\begin{equation}
\frac{K_0}{K_1} \log y =    \left[ \frac{\Gamma}{\Gamma-1} \log \left( K_1 \, y \right) \right].
\end{equation}
Moving all the terms with $\Gamma$ to one side and rearranging them, 
we obtain a functional form for the radial dependence of $\Gamma$:
\begin{eqnarray}
\Gamma(r) & = & \left( 1 - \frac{K_1}{K_0} \frac{\log (K_1 \, y)}{\log (y)} \right)^{-1},
\label{eq:gamma_r}
\end{eqnarray}
where $K_1 \sim 1.08$, and $K_0 \sim 4.36$ (see Fig.~\ref{fig:post_joint} for the posterior distributions, and relative covariance, of these parameters).
This functional form permits to model the radial variation of $\Gamma$, as shown in Fig.~\ref{fig:poly_r}.

\section{From the gas density to the hydrostatic mass profile}
\label{app:mass}

The hydrostatic equilibrium equation (Eq.~\ref{eq:HEE}) reads
\begin{equation}
\frac{1}{\mu m_p n_e} \frac{dP_e}{dR} = - \frac{G M(<R)}{R^2}.
\end{equation}

The term on the left side can be re-written in the following way:
\begin{equation}
\frac{1}{n_e} \frac{d P}{d R} = \frac{T}{R} \frac{d \log P}{d \log R} 
= \frac{T}{T_{500}} \frac{T_{500}}{R} 
\frac{d \log P/P_{500}}{d \log R/R_{500}},
\end{equation}
with $T_{500} = (G M_{500} \mu m_p) / (2 R_{500} )$ \citep[e.g.][]{voit+05}.

Using now Eq.~\ref{eq:gamma_Tr}
\begin{equation}
\frac{d \log P/P_{500}}{d \log R/R_{500}} = (\Gamma_0 + 1) \frac{d \log \left[ n_e E^{-2}(z) \right]}{d \log x} + \Gamma_R,
\end{equation}
with $x = R/R_{500}$, and adopting a NFW mass model, 
$M_{NFW} = M_{500} \mu_{NFW}(x)$ 
with $\mu_{NFW}(x) = \frac{\log (1 + c_{500} x) - c_{500} x/(1 + c_{500} x)}{\log (1 +  x) - x/(1 + x)}$), we can finally write 
the hydrostatic equilibrium equation as
\begin{equation}
T_0 \left[ n_e E^{-2}(z) \right]^{\Gamma_0} x^{\Gamma_R} \left( (\Gamma_0 + 1) \frac{d \log \left[ n_e E^{-2}(z) \right]}{d \log x} + \Gamma_R \right) = -\frac{2 \mu_{NFW}(x)}{x}. 
\end{equation}

If we redefine 
\begin{equation}
f(x) = \left[ n_e E^{-2}(z) \right]^{\Gamma_0},
\label{eq:redef}
\end{equation}
we can write 
\begin{equation}
T_0 f(x) x^{\Gamma_R} \left( \frac{\Gamma_0 + 1}{\Gamma_0} \frac{d \log f(x)}{d \log x} + \Gamma_R \right) = -\frac{2 \mu_{NFW}(x)}{x}. 
\end{equation}

Rearranging the terms in the last equation, we obtain
\begin{equation}
\frac{\Gamma_0 + 1}{\Gamma_0} \frac{d  f(x)}{d  x} x^{\Gamma_R} + {\Gamma_R} x^{\Gamma_R-1} f(x) = -\frac{2 \mu_{NFW}(x)}{T_0 x^2} 
\end{equation}
that implies
\begin{equation}
f'(x) + \frac{{\Gamma_R}}{x} \frac{\Gamma_0}{\Gamma_0 +1} f(x) = -\frac{2 \mu_{NFW}(x)}{T_0 x^2 x^{\Gamma_R}} \frac{\Gamma_0}{\Gamma_0 +1}.
\end{equation}
This is a partial differential equation (PDE) which can be easily solved once two functions $p(x)$ and $h(x)$ are defined as
\begin{equation}
p(x) = \frac{\Gamma_R}{x} \frac{\Gamma_0}{\Gamma_0 +1} \\
h(x) = -\frac{2 \mu_{NFW}(x)}{T_0 x^2 x^{\Gamma_R}} \frac{\Gamma_0}{\Gamma_0 +1},
\end{equation}
and the PDE can be simplified to
\begin{equation}
f' + p f = h.
\end{equation}

Then by computing a new function
\begin{equation}
\nu(x) = \exp \left[ \int p(t) dt + K \right],
\end{equation}
we can finally solve for 
\begin{equation}
f(x) = \frac{\int \nu(t) h(t) dt + c}{\mu(x)}.
\end{equation}

In summary: given a set of the two NFW parameters, we can estimate $f(x)$, and, by inverting Eq.~\ref{eq:redef}, we obtain a model for the density profile.
Then, by applying a maximum likelihood technique to minimize the distance between the model of the gas density 
and the observed profile $n_e(x)$, we can find the best fitting NFW parameters, and thus measure directly $M_{500}$ and $c_{500}$ as well as the entire mass profile.

\end{appendix}
\end{document}